\makeatletter
\@ifundefined{@parse@version@dash}{%
\def\@parse@version#1{\@parse@version@0#1}
\def\@parse@version@#1/#2/#3#4#5\@nil{%
\@parse@version@dash#1-#2-#3#4\@nil}
\def\@parse@version@dash#1-#2-#3#4#5\@nil{%
  \if\relax#2\relax\else#1\fi#2#3#4 }
}{}
\makeatother
\documentclass[prl,twocolumn,superscriptaddress]{revtex4-2}
\usepackage[dvipsnames]{xcolor}
\usepackage{amsmath}
\usepackage{units,xspace}
\usepackage{graphicx}
\usepackage{array}
\usepackage{lineno}
\usepackage{url,hyperref}
\usepackage{multirow}
\usepackage[utf8]{inputenc}
\newcommand{\ifb}{\mbox{fb$^{-1}$}\xspace}

\newcommand{\tev}{\mbox{TeV}\xspace}
\newcommand{\gev}{\mbox{GeV}\xspace}

\newcommand{\gd}{\ensuremath{g_\textrm{D}}\xspace}

\newcommand{\MAD}{\textsc{MadGraph5}\xspace}

\newcommand{\hf}{\nicefrac{1}{2}\xspace}

\DeclareGraphicsExtensions{.pdf,.png,.jpg,.jpeg}
\graphicspath{{./}{./fig/}}

\begin{document}


\sloppy




\title{\vspace{-2.0cm} Search for Highly-Ionizing Particles in  \\ $pp$ Collisions During LHC Run-2 Using the Full MoEDAL Detector}

\author{B.~Acharya}
\altaffiliation[Also at ]{Int. Centre for Theoretical Physics, Trieste, Italy}   
\affiliation{Theoretical Particle Physics \& Cosmology Group, Physics Dept., King's College London, UK}

\author{J.~Alexandre}
\affiliation{Theoretical Particle Physics \& Cosmology Group, Physics Dept., King's College London, UK}

\author{P.~Benes}
\affiliation{IEAP, Czech Technical University in Prague, Czech~Republic}

\author{B.~Bergmann}
\affiliation{IEAP, Czech Technical University in Prague, Czech~Republic}

\author{S.~Bertolucci}
\affiliation{INFN, Section of Bologna, Bologna, Italy}

\author{A.~Bevan}
\affiliation{School of Physics and Astronomy, Queen Mary University of London, UK}

\author{R.~Brancaccio}
\affiliation{INFN, Section of Bologna \& Department of Physics \& Astronomy, University of Bologna, Italy}

\author{H.~Branzas}
\affiliation{Institute of Space Science, Bucharest - M\u{a}gurele, Romania} 

\author{P.~Burian}
\affiliation{IEAP, Czech Technical University in Prague, Czech~Republic}

\author{M.~Campbell}
\affiliation{Experimental Physics Department, CERN, Geneva, Switzerland}

\author{S.~Cecchini}
\affiliation{INFN, Section of Bologna, Bologna, Italy}

\author{Y.~M.~Cho}
\affiliation{Center for Quantum Spacetime, Sogang University, Seoul, Korea} 

\author{M.~de~Montigny}
\affiliation{Physics Department, University of Alberta, Edmonton, Alberta, Canada}

\author{A.~De~Roeck}
\affiliation{Experimental Physics Department, CERN, Geneva, Switzerland}

\author{J.~R.~Ellis}
\affiliation{Theoretical Particle Physics \& Cosmology Group, Physics Dept., King's College London, UK}
\affiliation{Theoretical Physics Department, CERN, Geneva, Switzerland}


\author{M.~Fairbairn}
\affiliation{Theoretical Particle Physics \& Cosmology Group, Physics Dept., King's College London, UK}

\author{D.~Felea}
\affiliation{Institute of Space Science, Bucharest - M\u{a}gurele, Romania}

\author{M.~Frank}
\affiliation{Department of Physics, Concordia University, Montr\'{e}al, Qu\'{e}bec,  Canada}

\author{J.~Hays}
\affiliation{School of Physics and Astronomy, Queen Mary University of London, UK}

\author{A.~M.~Hirt}
\affiliation{Department of Earth Sciences, Swiss Federal Institute of Technology, Zurich, Switzerland}

\author{D.~L.-J.~Ho}
\affiliation{Department of Physics, Imperial College London, UK}

\author{P.~Q.~Hung}
\affiliation{Department of Physics, University of Virginia, Charlottesville, Virginia, USA}

\author{J.~Janecek}
\affiliation{IEAP, Czech Technical University in Prague, Czech~Republic}

\author{M.~Kalliokoski}
\affiliation{Helsinki Institute of Physics, University of Helsinki, Helsinki, Finland}



\author{D.~H.~Lacarr\`ere}
\affiliation{Experimental Physics Department, CERN, Geneva, Switzerland}


\author{C.~Leroy}
\affiliation{D\'{e}partement de Physique, Universit\'{e} de Montr\'{e}al, Qu\'{e}bec, Canada}

\author{G.~Levi} 
\affiliation{INFN, Section of Bologna \& Department of Physics \& Astronomy, University of Bologna, Italy}

\author{A.~Margiotta}
\affiliation{INFN, Section of Bologna \& Department of Physics \& Astronomy, University of Bologna, Italy}

\author{R.~Maselek}
\affiliation{Laboratoire de Physique Subatomique et de Cosmologie, Universit\'e Grenoble-Alpes CNRS/IN2p3, 
Grenoble, France}
\affiliation{Institute of Theoretical Physics, University of Warsaw, Warsaw, Poland}

\author{A.~Maulik}
\affiliation{INFN, Section of Bologna, Bologna, Italy}
\affiliation{Physics Department, University of Alberta, Edmonton, Alberta, Canada}

\author{N.~Mauri}
\affiliation{INFN, Section of Bologna \& Department of Physics \& Astronomy, University of Bologna, Italy}

\author{N.~E.~Mavromatos}
\altaffiliation[Also at ]{Department of Physics, School of Applied Mathematical and Physical Sciences, 
National Technical University of Athens, Athens, Greece} 
\affiliation{Theoretical Particle Physics \& Cosmology Group, Physics Dept., King's College London, UK}

\author{M.~Mieskolainen}
\affiliation{Physics Department, University of Helsinki, Helsinki, Finland}

\author{L.~Millward}
\affiliation{School of Physics and Astronomy, Queen Mary University of London, UK}

\author{V.~A.~Mitsou}
\altaffiliation[Also at ]{Department of Physics, School of Applied Mathematical and Physical Sciences, 
National Technical University of Athens, Athens, Greece} 
\affiliation{IFIC, CSIC -- Universitat de Val\`{e}ncia, Valencia, Spain}

\author{A. Mukhopadhyay}
\affiliation{Physics Department, University of Alberta, Edmonton, Alberta, Canada}

\author{E.~Musumeci}
\affiliation{IFIC, CSIC -- Universitat de Val\`{e}ncia, Valencia, Spain}


\author{I.~Ostrovskiy}
\affiliation{Department of Physics and Astronomy, University of Alabama, Tuscaloosa, Alabama, USA}

\author{P.-P.~Ouimet}
\affiliation{Physics Department, University of Regina, Regina, Saskatchewan, Canada}
  
\author{J.~Papavassiliou}
\affiliation{IFIC, CSIC -- Universitat de Val\`{e}ncia, Valencia, Spain}


\author{L.~Patrizii}
\email[Corresponding author: ]{Laura.Patrizii@bo.infn.it}
\affiliation{INFN, Section of Bologna, Bologna, Italy}

\author{G.~E.~P\u{a}v\u{a}la\c{s}}
\affiliation{Institute of Space Science, Bucharest - M\u{a}gurele, Romania}

\author{J.~L.~Pinfold}
\email[Corresponding author: ]{jpinfold@ualberta.ca}
\affiliation{Physics Department, University of Alberta, Edmonton, Alberta, Canada}

\author{L.~A.~Popa}
\affiliation{Institute of Space Science, Bucharest - M\u{a}gurele, Romania}

\author{V.~Popa}
\affiliation{Institute of Space Science, Bucharest - M\u{a}gurele, Romania}

\author{M.~Pozzato}
\affiliation{INFN, Section of Bologna, Bologna, Italy}

\author{S.~Pospisil}
\affiliation{IEAP, Czech Technical University in Prague, Czech~Republic}

\author{A.~Rajantie}
\affiliation{Department of Physics, Imperial College London, UK}

\author{R.~Ruiz~de~Austri}
\affiliation{IFIC, CSIC -- Universitat de Val\`{e}ncia, Valencia, Spain}

\author{Z.~Sahnoun}
\affiliation{INFN, Section of Bologna \& Department of Physics \& Astronomy, University of Bologna, Italy}

\author{M.~Sakellariadou}
\affiliation{Theoretical Particle Physics \& Cosmology Group, Physics Dept., King's College London, UK}

\author{K.~Sakurai}
\affiliation{Institute of Theoretical Physics, University of Warsaw, Warsaw, Poland}


\author{S.~Sarkar}
\affiliation{Theoretical Particle Physics \& Cosmology Group, Physics Dept., King's College London, UK}

\author{G.~Semenoff}
\affiliation{Department of Physics, University of British Columbia, Vancouver, British Columbia, Canada}

\author{A.~Shaa}
\affiliation{Physics Department, University of Alberta, Edmonton, Alberta, Canada}

\author{G.~Sirri}
\affiliation{INFN, Section of Bologna, Bologna, Italy}

\author{K.~Sliwa}
\affiliation{Department of Physics and Astronomy, Tufts University, Medford, Massachusetts, USA}

\author{R.~Soluk}
\affiliation{Physics Department, University of Alberta, Edmonton, Alberta, Canada}

\author{M.~Spurio}
\affiliation{INFN, Section of Bologna \& Department of Physics \& Astronomy, University of Bologna, Italy}

\author{M.~Staelens}
\affiliation{IFIC, CSIC -- Universitat de Val\`{e}ncia, Valencia, Spain}

\author{M.~Suk}
\affiliation{IEAP, Czech Technical University in Prague, Czech~Republic}

\author{M.~Tenti}
\affiliation{INFN, Section of Bologna, Bologna, Italy}

\author{V.~Togo}
\affiliation{INFN, Section of Bologna, Bologna, Italy}

\author{J.~A.~Tuszy\'{n}ski}
\affiliation{Physics Department, University of Alberta, Edmonton, Alberta, Canada}

\author{A.~Upreti}
\affiliation{Department of Physics and Astronomy, University of Alabama, Tuscaloosa, Alabama, USA}

\author{V.~Vento}
\affiliation{IFIC, CSIC -- Universitat de Val\`{e}ncia, Valencia, Spain}

\author{O.~Vives}
\affiliation{IFIC, CSIC -- Universitat de Val\`{e}ncia, Valencia, Spain} 

\collaboration{THE MoEDAL COLLABORATION}
\noaffiliation

\vspace*{1.0cm}


\begin{abstract} This search for 
Magnetic Monopoles (MMs) and High Electric
Charge Objects (HECOs) with spins 0, 1/2 and 1, 
uses for the first time the full MoEDAL detector, exposed to 6.46~fb$^{-1}$ proton-proton collisions at 13 TeV.  The results are interpreted in terms of Drell-Yan and photon-fusion pair production. Mass limits on direct production of MMs of up to 10 Dirac magnetic charges and HECOs with electric charge in the range 10$e$ to 400$e$, were achieved. The charge limits placed on MM and HECO production are currently the strongest in the world.  MoEDAL is the only LHC experiment capable of being directly calibrated for highly ionizing particles using heavy ions and with a detector system dedicated to definitively measuring magnetic charge.
\end{abstract}


\maketitle


The detection of a massive, stable or pseudo-stable,  Highly-Ionizing Particle (HIP) with a significantly large electric charge $|q| \gg e $ (where $e$ is the elementary charge) and/or a magnetic charge $\gd$, would provide compelling evidence for physics beyond the Standard Model. Many hypothetical particles capable of producing HIP signatures have been proposed, including Magnetic Monopoles (MMs)~\cite{Dirac:1931,thooft:1974,polyakov:1974,Cho:1997,Cho:2005,CKY:2015,hungmm,EMY:2016,Fairbairn:2007,Giacomelli:2011,Patrizii-Spurio:2015,Mavromatos:2020gwk} and dyons~\cite{Schwinger:1969};  Q-balls  \cite{Coleman85,Kusenko98};  micro black-hole remnants~\cite{Koch07}; doubly charged  massive particles~\cite{Acharya:2014}; scalars in neutrino-mass models~\cite{Hirsch:2021}; and aggregates of $ud$-~\cite{Holdom:2018} or $s$-quark matter~\cite{Farhi84}.

HIPs have been sought in matter~\cite{Burdin:2015}, cosmic rays \cite{Burdin:2015,Patrizii-Sahnoun:2019} and in accelerator-based experiments
\cite{Fairbairn:2007,Pinfold:2009,Mavromatos:2020gwk}, with recent searches conducted at the LHC \cite{Aad:2011,Aad:2012,Aad:2013,Chatrchyan:2013,Aad:2015,Aad:2016,Acharya:2016,Acharya:2017,Acharya:2018,Acharya:2019,Aaboud:2019,MoEDAL:2020pyb,MoEDAL:2021vix,Acharya:2022,ATLAS-13TeV,ATLAS:2023zxo,Aad:2023}.

According to the Dirac Quantization Condition \cite{Dirac:1931}, magnetic charge would be quantized in units of  $\gd = e/2\alpha \approx 68.5e$, where $\alpha$ is the fine structure constant. Due to the exceptionally strong coupling between a MM and the photon, perturbative calculations cannot reliably predict cross-sections associated with MM production. 
The diagrams involved need to undergo appropriate resummation in order to account for potential non-perturbative quantum corrections \cite{Roberts:1994dr,Binosi:2009qm,Alexandre:2019}. Such techniques have been investigated, especially for High Electric Charge Objects (HECOs)~\cite{Alexandre:2023qjo}. For simplicity and ease of comparison with other search results, here we use leading-order Feynman diagrams for cross-section estimates.

  \begin{figure}[h]
    \centering
    \includegraphics [width=0.8\linewidth]{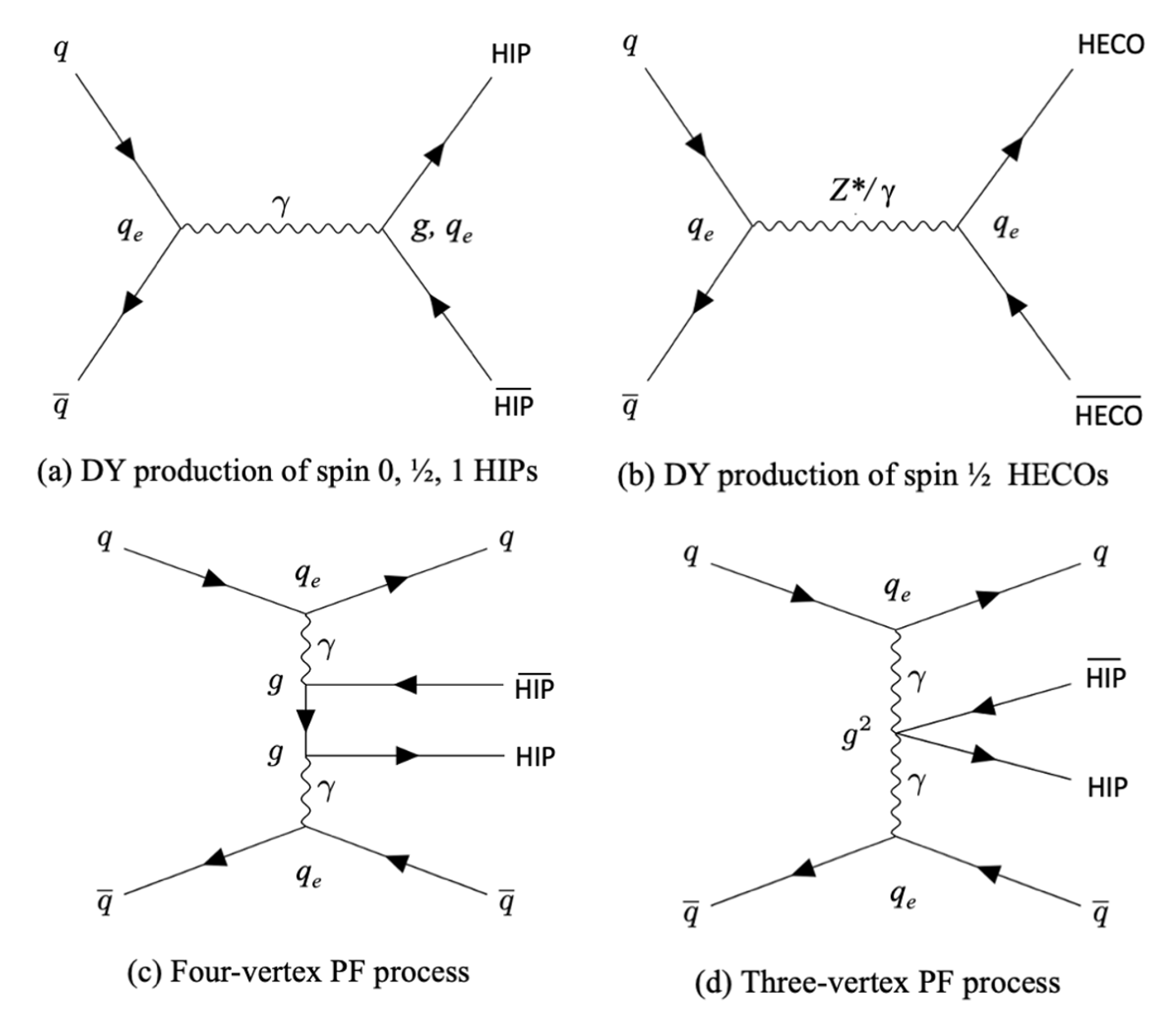}
    \caption{Tree-level Feynman diagrams for the following:  (a) DY  production of HIP--anti-HIP pairs; (b)  spin-\hf
    HECO pairs; and, (c,d)  PF production of HIP--anti-HIP pairs.}
    \label{fig:DY-HIPs}
\end{figure}
  
 The Drell--Yan (DY) pair production mechanism provides a basic model for HIP creation. HIP pair production of  spin-0, spin-1/2 and spin-1 are computed using the Feynman diagrams shown at the top of  Fig.~\ref{fig:DY-HIPs}. Spin-\hf  HECO DY production can take place via virtual photon ($\gamma$) or $\gamma/Z^{0}$ exchange~\cite{Wendy}. For DY production of spin-1/2 MMs,  the coupling of the $Z$ boson to magnetic charge is conventionally assumed to be absent. In some MM models, this can be proven explicitly~\cite{hungmm,mavromatosmm}.


 Additionally, MoEDAL has pioneered the search for HIPs at the LHC  using  photon fusion (PF)
 \cite{Baines-2018,Acharya:2019,Andreev:2005,Abbott:1998} described by the two Feynman diagrams shown at the bottom of  Fig.~\ref{fig:DY-HIPs}. For the mass range considered here, PF production of HIPs has a considerable cross-sectional advantage compared to  DY production~\cite{Baines-2018}. Consequently,  PF dominates when setting mass bounds. This picture is maintained even after resummation \cite{Alexandre:2023qjo}.

The physics processes considered here were generated using \MAD~\cite{Alwall:2014} utilizing the Universal FeynRules Output presented in Ref.~\cite{Baines-2018}.
The parton distribution functions (PDFs) \texttt{NNPDF23}~\cite{NNPDF:2012} and \texttt{LUXqed}~\cite{Manohar:2016} were used for the DY and PF production processes, respectively.  The \texttt{LUXqed} PDF was created in a model-independent way utilizing $ep$ scattering data.

In  this search
we used, for the first time, the full Run-2  MoEDAL detector.
MoEDAL's detector technology differs radically from the general-purpose LHC  experiments, ATLAS and CMS. The MoEDAL detector,     positioned around LHCb's VErtex LOcator detector (VELO) at IP8, employs two unconventional passive detection methodologies dedicated to HIP detection.
The first is a system of plastic Nuclear Track Detectors (NTDs) --- including the High-Charge Catcher (HCC) ---
 designed to register HIP ionization trails. The second detector system, the Magnetic Monopole Trapper (MMT), is designed to capture HIPs that stop within its sensitive volume, allowing for further study at the ETH Zurich SQUID magnetometer facility.

Both NTD and MMT detector systems require neither readout electronics nor a trigger since
no Standard Model (SM) particles can produce the distinctive signatures of  HIPs traversing the MoEDAL detector.  Thus, only a few HIP  messengers of new physics observed in MoEDAL data are needed to herald a discovery.
 Below we give a brief description of the MoEDAL detector.  A more detailed description of the detectors and their calibration and analysis is given in the supplemental material~\cite{extra_material}. 

The NTD detector system is positioned around LHCb's VELO detector. 
It consists of an array of standard NTD stacks and the HCC, both of which are comprised of Makrofol NTD plastic.
The HCC array is designed to extend the search for  HIPs to the highest charge possible. This is achieved by situating the HCC between LHCb's  RICH1 detector and the first LHCb downstream tracking detector, TT1, minimizing the material in which the very highest ionizing particles may be absorbed. After exposure in the LHC's IP8 region, the MoEDAL NTD stacks are sent to  INFN Bologna for etching and scanning.  

The pertinent quantity for MoEDAL's plastic NTDs is the Restricted Energy Loss (REL) \cite{REL}.
Heavy ion beams are used to determine the NTD response over a wide range of energy loss, as explained in Ref.~\cite{calib2007}. The NTD  response  is  calibrated directly using  heavy-ion beams  at the NASA Space Radiation Facility (NSRL) at the Brookhaven National Laboratory in the US and at the CERN SPS. 
The REL corresponding to the threshold of the Makrofol NTDs utilized is $\mathrm{\sim 2700~MeV\,g^{-1}cm^{2}}$.

A study of the HIP detection efficiency of NTDs in the presence of beam backgrounds was performed by using  NTD  calibration stacks exposed to a relativistic lead-ion beam.  The stacks were comprised of  Makrofol NTD foils, exposed to beam backgrounds as part of MoEDAL's NTD array at LHC's Run-2 during 2018, 
interleaved with previously $\it{unexposed}$
 Makrofol NTD sheets using plastic from the same production batch 
 utilized  in calibration and standard data taking.
The  NTDs sheets comprising the calibration stacks were then etched 
and scanned using the same techniques and procedures 
employed to examine all MoEDAL NTD stacks. As the relativistic lead-ion calibration beam particles penetrate the whole stack, 
the signal etch-pits seen in the LHC unexposed sheets can serve as a map of the passage of the heavy-ions.
The identification efficiency of etch-pits in the LHC-unexposed sheets is measured to be effectively 100\% by making independent comparison scans of the other LHC-unexposed sheets in the stack which
have the identical etch-pit number and pattern.

 

During Run-2 the surface area of standard NTD stacks facing IP8 was 10.7 m$^{2}$. The corresponding area of HCC foils was 3.24 m$^{2}$. To avoid excessive damage from beam-induced backgrounds the standard stacks were replaced three times. The HCC foils were replaced more frequently, 7 times in all,  as they were deployed nearer the beam line.

A pair of collinear incident and exiting etch-pits, found in the first layer of a 6-layer NTD stack,
 is defined as a  ``candidate''. 
The discovery of such a candidate would trigger the analysis of all 5 downstream foils comprising an NTD stack.
A HIP ``candidate track'' requires collinear etch-pits pairs,  pointing to the collision point, in all six NTD sheets in the stack. No ``candidate'' or  ``candidate track'' was found.

The MMT detector comprises 2400 aluminium bars with a mass of 800 kg deployed in three arrays transverse to and just forward of IP8. Aluminium was chosen as the trapping material because the aluminium nucleus has an anomalously large magnetic moment which engenders a monopole--nucleus binding energy of 0.5--2.5~MeV~\cite{binding},  which is consonant with the shell-model splittings.
A MM will stop in the MMT if its speed falls below  $\beta \le 10^{-3}$. It will then bind to the nucleus due to the interaction between the MM and the nuclear magnetic moment~\cite{Milton2006,binding,goebel:1984,bracci:1984,olaussen:1985}.
MMs bound in such a way would be trapped indefinitely~\cite{Milton2006,binding}.

After the MMT's aluminium volumes were exposed for roughly a year, they were sent to the ETH Zurich Laboratory for Natural Magnetism. They were passed through a SQUID magnetometer to check for trapped magnetic charges.The SQUID magnetometer calibration was performed using a known magnetic dipole sample and checked using a simulated MM comprised of long thin solenoids, which provided the same response as a MM with a known magnetic charge.

 A potential MMT inefficiency from charge nullification arises if a monopole and an anti-monopole are trapped in the same volume. With no monopoles detected in Run-2 we only consider the case where, at most, a few MMs are potentially detectable in the MMT. However, with 2400 MMT volumes available this effect on MMT efficiency is deemed negligible.


 \begin{figure*}[htb]
  \begin{center}
  \includegraphics[width=0.49\linewidth]{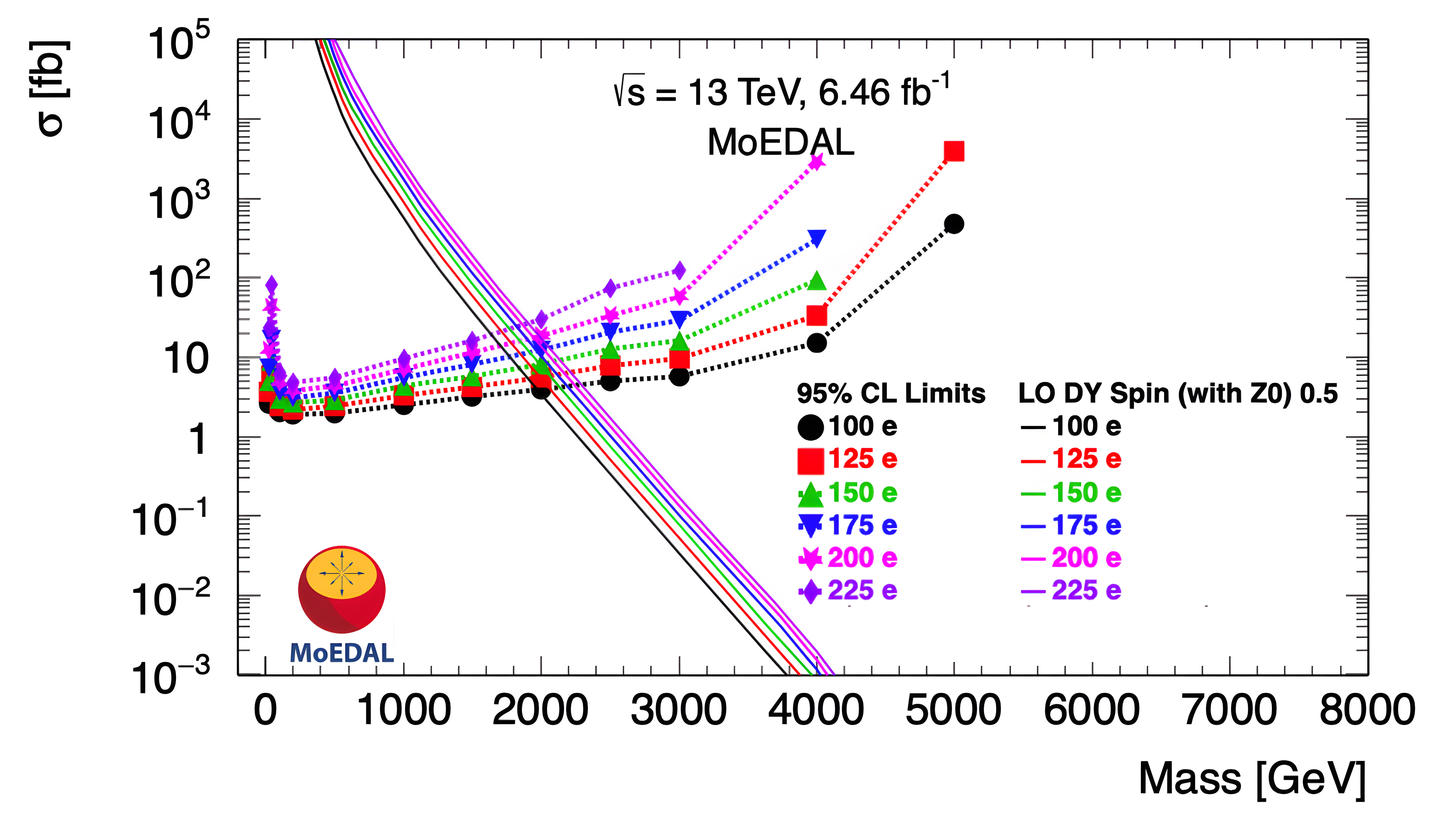} 
   \includegraphics[width=0.49\linewidth]{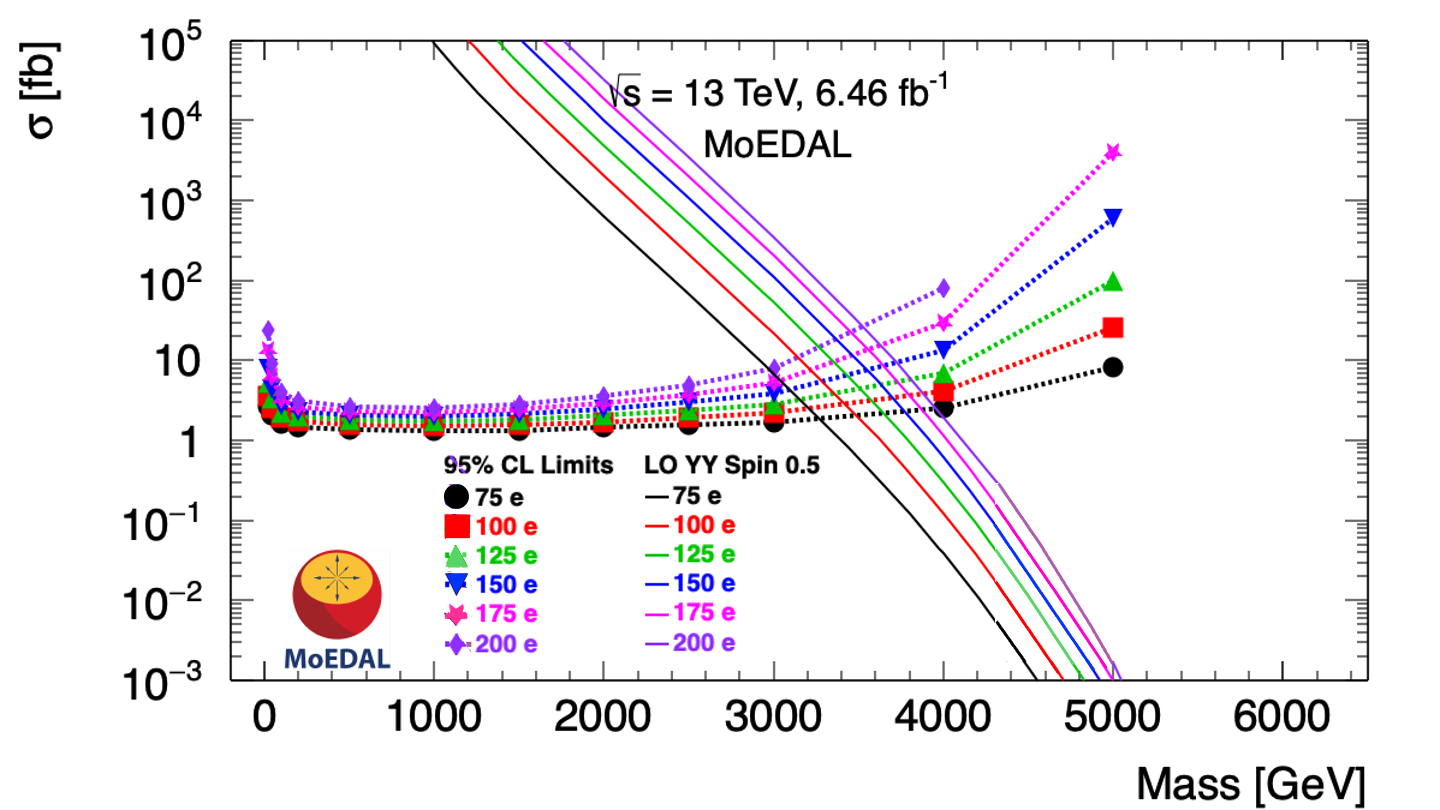}
  \includegraphics[width=0.49\linewidth]{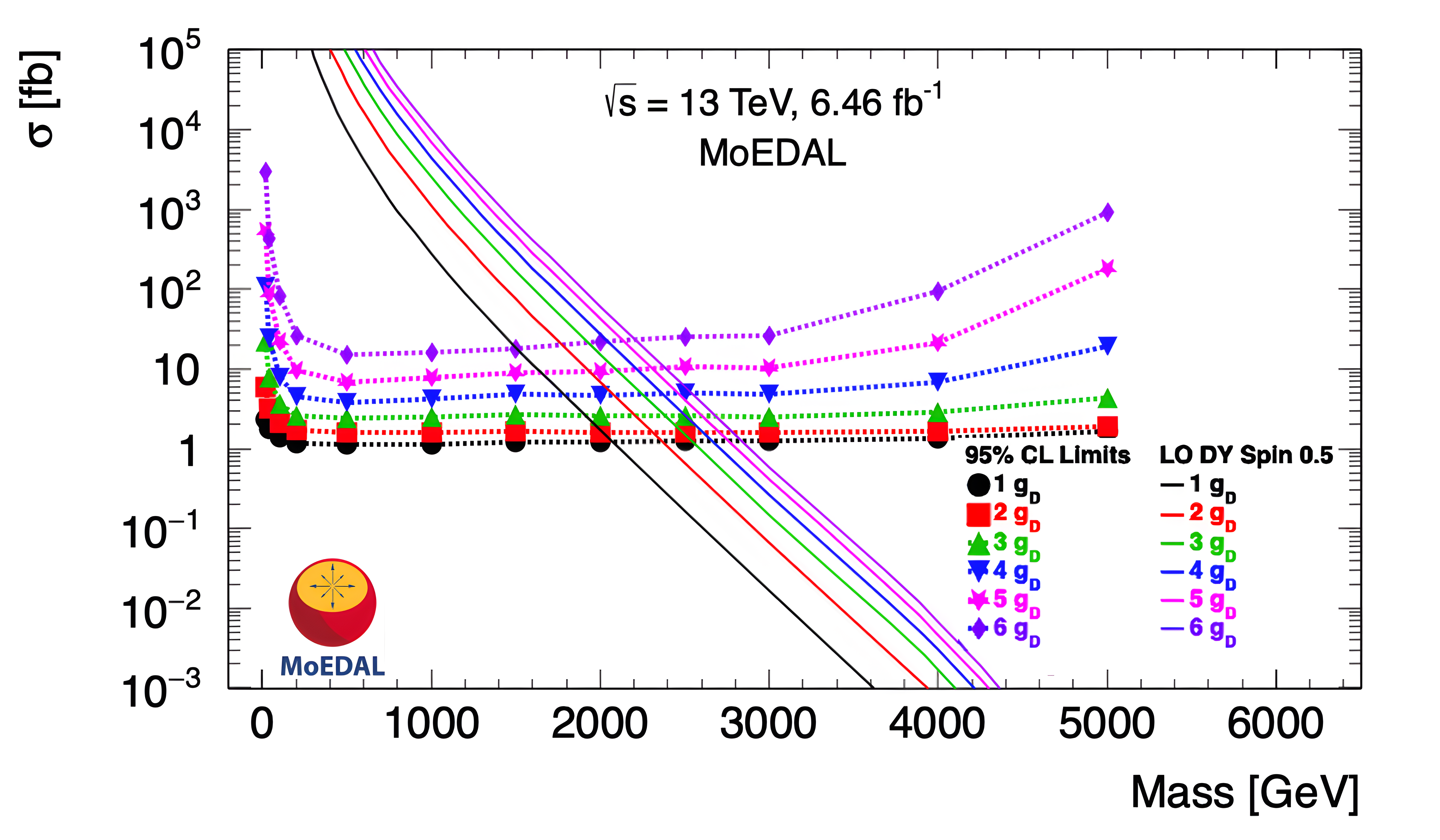} 
  \includegraphics[width=0.49\linewidth]{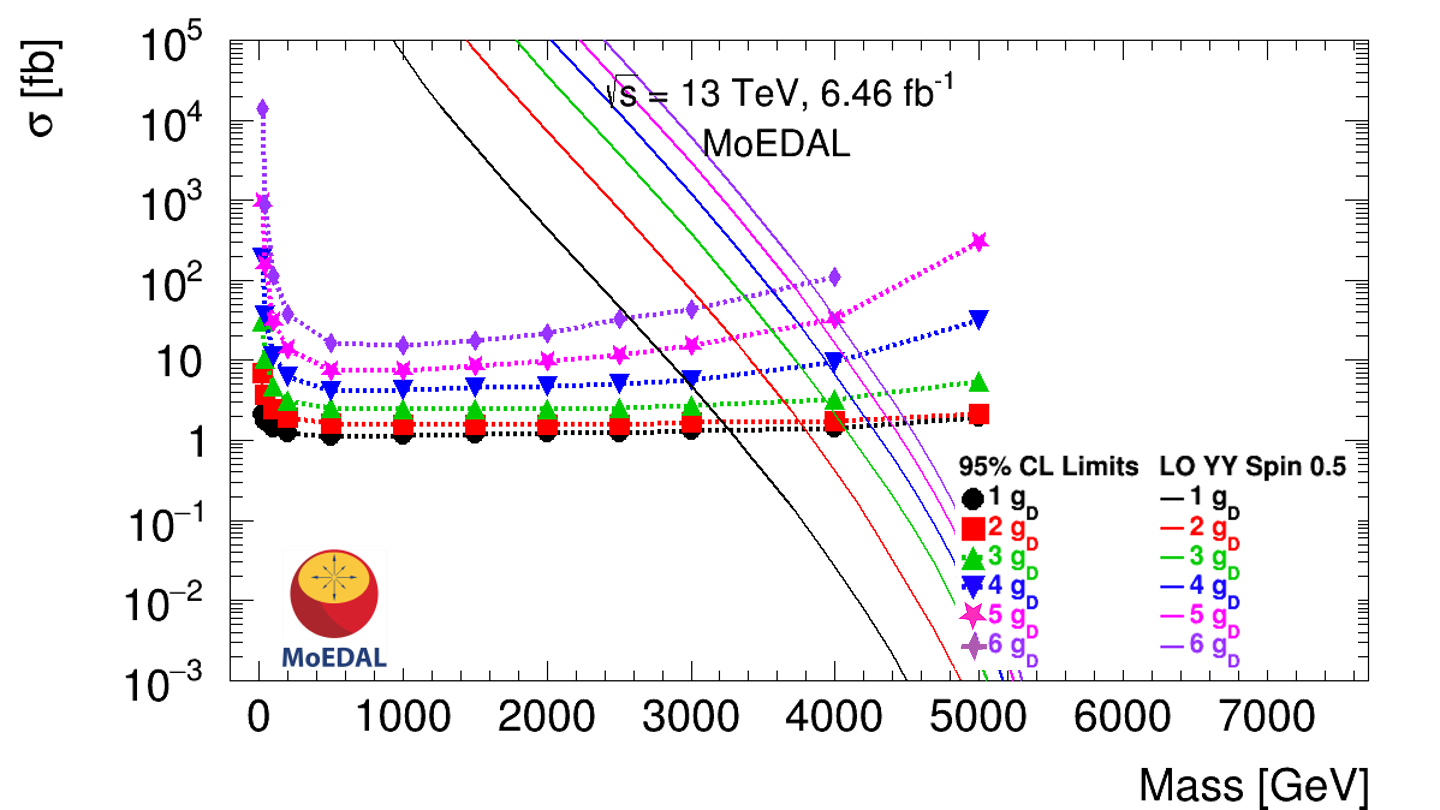}  
 \end{center}
  \caption{\label{fig:limits1} 95\% C.L.\ upper limits to the cross-section for the following production mechanisms: (Top Left) 
   a DY model with virtual $\gamma/Z^{0}$ exchange for spin-\hf production of HECOs with charge in the range $100e$--$225e$; 
   (Top Right) PF production of HECOs with charge $q$ in the range $100e$--$225e$;  
   (Bottom Left)  a DY model  with virtual photon spin-\hf production of MMs with magnetic charge in the range $1\gd$--$6\gd$; and 
    (Bottom Right)  PF production of MMs with magnetic charge in the range $1\gd$--$6\gd$.
   The solid lines denote the cross-section predictions for each case considered.}
\end{figure*}


   \begin{table*}[htb]\centering
  \footnotesize
\caption{\label{tab:masslimits-monopoles} 95\% CL lower mass limits for the MM  search. For  14  high-charge  entries,  a range of masses has been excluded }
\begin{tabular}{|c|c|cccccccccc|} \hline
    &     & \multicolumn{10}{c|}{Magnetic charge (\gd)} \\
Spin  & Process   & 1        &        2   &   3       &   4        &  5      &   6      &  7     &  8      &  9   & 10     \\ 
\cline{3-12}
& & \multicolumn{10}{c|}{95\% CL mass limits (\gev)} \\
\colrule
0  & DY                  & 1450 & 1660  & 1730  & 1680  & 1590 & 1510 & 1380 & 100--1210 & 200--980 & 500--790  \\
\hf & DY                  & 2070 & 2300  & 2370 & 2360  & 2300 & 2200 & 2030 & 100--1810 & 100--1470& 500--1040  \\
1 & DY                     & 2180 & 410  & 2510 & 2520  & 2460 & 2370 & 2240 & 2090 & 1870 & 200--1550 \\
 \hline
 0 & $\gamma\gamma$  & 3010 & 3510 & 3700 & 3730& 3680  & 3550 & 3370 & 100--3000  & 1000--2000  & .... \\
\hf & $\gamma\gamma$ & 3240 & 3730 & 3920 & 3940& 3880 & 3800 & 3590  & 3000  & 100--2500  & 500-1500 \\
1 & $\gamma\gamma$   & 3650 & 4100 & 4220 & 4230& 4170& 4080 & 3960   & 100--3500  & 100--3000 & 500-2000  \\
 \hline
\end{tabular}
\end{table*}

\begin{table*}[htb]
\footnotesize
\caption{\label{tab:mass-limits-hecos} 95\% CL lower mass limits for the HECO search. For 2 entries, at 10$e$ and 400$e$,  a range of masses has been excluded}
\begin{tabular}{|c|c|ccccccccccccccccccc|} \hline
   &  \multirow{3}{1.5cm}{Process/ exchange}  & \multicolumn{19}{c|}{Electric charge ($e$)}  \\
Spin  &     &  10  & 15    &  20   &  25    &  50    & 75     &  100  &  125  & 150  &  175   & 200   & 225  & 250 &  275  &  300  & 325  & 350 & 375 & 400 \\ 
\cline{3-21}
&   & \multicolumn{19}{c|}{95\% CL mass limits (\gev)  } \\
\colrule
0  & DY $\gamma$               & 80  & 220 & 400 & 580 & 1300&1390 &1420& 1430&1430& 1410&1390&1370&1340&1290&1210&1070& 900 & 500   & .... \\
\hf & DY $\gamma$             & 290& 620 & 920 &1190& 1850&1940 & 1980&2000&2000& 1980&1940&1900&1830&1740&1610&1380 & 1000 &  200   & .... \\
\hf & DY $\gamma/Z^{*}$  & 320& 620 & 930 &1170& 1840& 1930 &1970&1990&1980& 1970&1940&1900&1840&1740& 1620& ....  & .... &  .... & .... \\
1  & DY $\gamma$               & 200--330& 640 & 960 &1240& 2020& 2120&2170& 2180& 2180&2170&2140&2100&2060&1980& 1850&1620& 1000 & 100  & ....  \\  
\hline
0  & $\gamma\gamma$   & 300&740 &1140&1500& 2720&3020 &3170& 3260&3310& 3320&3310&3240&3160& 3000 & 2500  & 2000   &  1500 &  1500 & .... \\
\hf & $\gamma\gamma$  &410 & 950 &1380&1800&3000 &3250 &3390& 3480&3520& 3530&3500&3440& 3000 & 3000 & 2500 &  2000  & 1500  & 1500   & 500--1500 \\
1  & $\gamma\gamma$   & 790&1400&1880&2310&3400 &3640 &3770& 3850&3890& 3890&3850&3730& 3000 & 3000 & 2500  & 2000  &  1500  &  1500  & .... \\   \hline
\end{tabular}
\end{table*}
 





The acceptance of the MoEDAL detector at Run-2 is defined to be the fraction of events in which at least one HIP of the produced pair was detected in MoEDAL in either the NTD detector or the MMT detector,
giving two sets of acceptance curves, one for each production mechanism considered.
The acceptance for HECOs and MMs depends on an interplay between the positions of MoEDAL detector modules; energy loss in the detectors; the mass and charge of the particles; and the spin-dependent kinematics of the interaction products. For HECOs, MoEDAL's NTD system provides the only means of detection. For charges $\le$4gD the NTD acceptance is roughly ten times that of the MMTs across the mass range. For higher charges, the MMT acceptance is 1\%, or less, of the NTD value. This story is essentially the same for DY MM production. 

The spin-dependent acceptance for
different spins are mostly due to the degree of coincidence of the azimuthal ($\phi$) and polar ($\theta$) distributions of the produced HIPs  with that
 of the detector elements.  Example acceptance curves 
 are provided in the supplemental material~\cite{extra_material}.
The  Run-2 geometric acceptance  of MoEDAL's NTD and MMT detectors for MMs
partly overlap. Thus, care must be taken not to double-count signals when calculating the acceptance of the overall detector.  Only the NTDs can be utilized for the HECO analysis since we cannot register HECOs
trapped in the MMT detectors.

The largest source of systematic uncertainty on the acceptance is due to the imperfect knowledge of the material between the VELO vacuum tank and the MoEDAL detector modules. It is described by three models of  MoEDAL's detector material. First,  the Minimal Model is estimated from the LHCb detector material map and the available material description external to  VELO's vacuum tank. The second, Nominal Model,  includes an additional layer of material amounting to 0.15 Radiation Lengths (RLs),  corresponding to our estimate for extra material such as cabling and other VELO-related infrastructure. The third, called the Maximal Model contains a conservative assessment of the amount of extra material amounting to 0.3 RLs. Estimates indicate the material lying between the VELO detector and the MoEDAL detector elements is on average 1.4 RLs  \cite{LHCb-detector} but can be as much as  8.0 RLs. The uncertainty in the acceptance using this description of the material is estimated by a GEANT4-based Monte Carlo simulation to lie in the range of  1\% to 25\%. The uncertainty is estimated for each combination of mass and charge studied.

The energy loss for MMs and HECOS is estimated over the range of velocities,  $10^{-4}  \le \beta \le 1$. The relative uncertainty on  $dE/dx$ is estimated to be  10\% for $\beta \ge $ 0.1  (High-$\beta$) and for $  \beta  <  0.01$  (low-$\beta$)  30\%  \cite{Acharya:2016}. The $dE/dx$ for $0.01 \le  \beta \le 0.1$ (mid-$\beta$) is determined by interpolation between the high-$\beta$ and low-$\beta$ regimes, the relative uncertainty in this velocity range is set according to a velocity-dependent linear extrapolation between 10\% and  30\%. A Monte Carlo study showed that the systematic uncertainty on the acceptance at high-$\beta$ is 1\% to 7\% and for  mid-$\beta$ between 1\% and 9\%. Conservatively,  we set the systematic uncertainty on the acceptance from the error on  $dE/dx$ to be 10\% for all velocities.

Another source of systematic uncertainty on the acceptance is due to the uncertainty in the positioning of the MMT and  NTD' Detectors. The NTD stacks and MMT array positioning errors are estimated to be as much as $\pm$0.5 cm and $\pm$1cm, respectively,  in each coordinate.  Monte Carlo studies show that the overall systematic uncertainty in the acceptance resulting is dominated by the NTD placement error. Monte Carlo studies show that the uncertainty in the positioning of the MMT and NTD detectors gives rise to a systematic uncertainty in the acceptance that ranges from 1\% to 8\%.
 A conservative error of $\pm$ 10\% is assigned from this source for all electric and magnetic charges.   
 
 In this analysis, the luminosity determination made by  LHCb  \cite{lhcb_lumi} was utilized. The error on the luminosity is assessed to be 3.9\% \cite{BGI}. This error is summed in quadrature with the other systematic uncertainties to be included in the determination of the 95\% CL limits reported here.
 
 To illustrate the effect of the systematic uncertainties on the 95\% CL limits, we consider these examples: MMs with charge 1\gd and spin-\hf  as well as   HECOs with charge 100e and spin-\hf in the mass range 20 GeV to 5 TeV. The fractional systematic uncertainty on the acceptance from all the sources discussed above is approximately 14\% for all MM masses. For HECOs this error is around 15\% over most of the range, finally increasing to around 30\% at 5 TeV. As shown in the Supplemental Material \cite{extra_material}, in both cases, the effect of this error on the 95\% CL contours was small and did not affect the 95\% mass limits.

No MM or HECO  candidates were found after exposure of the MoEDAL detector to  6.46~\ifb of $pp$ collisions at a center-of-mass energy of 13~\tev during LHC's Run-2.  Consequently, MoEDAL placed 95\% confidence level (CL) lower mass limits on  spin-1, spin-\hf and spin-0 MM and HECO production.  A complete set of limit curves for all HIP charge and mass points considered is included in the supplemental material~\cite{extra_material}.  No limits are set for a particular value of mass and charge if the acceptance is  $< $ 0.1\%  and the systematic uncertainties are   $\ge$100\%.
 
Examples of the limit curves obtained for spin-\hf  HECOs produced via  $\gamma/Z^{0}$ exchange and photon-fusion, with charges ranging from $100e$--$225e$, are given in Fig.~\ref{fig:limits1} (top left)  and (top right), respectively.
In the MM case, the example limit curves for MMs produced via DY and photon-fusion with charge from $1\gd$ to $6\gd$ are shown in  Fig.~\ref{fig:limits1} (bottom left)  and (bottom right), respectively.

The results of  MoEDAL's search for MMs at the LHC reached a
sensitivity to DY production cross-section with cross-section as small as  1~fb.
Mass limits as high as  $\sim2.5~\tev$ were placed on magnetic charges up to  10\gd.  Corresponding 95\% CL mass limits were placed on  MMs produced via photon-fusion for: i) cross-section as low as  1~fb ; ii)  magnetic charge reaching up to 10\gd; and iii) MM mass as high as  $\sim 4.2~\tev$. MoEDAL's lower mass limits are summarized in Table~\ref{tab:masslimits-monopoles}. A sample of the curves obtained for MMs produced by DY and photon-fusion is given in Fig.~\ref{fig:limits1} (bottom left)  and (bottom right), respectively.  A description of the behaviour of the limits as a function of mass for HECOs and MMs is provided in the supplemental material \cite{extra_material}.

Interpreting our results as a search for HECOs, we placed  95\% CL lower mass limits on the following: i)  HECO production via the DY mechanism for HECOs of spin-0, \hf and 1; ii) cross-sections as low as 2~fb ; iii) electric charges in the range $10e \le |q| \le 375e$; and iv) mass limits ranging up to 2.2~\tev. For spin-\hf HECOS, we also considered DY production via $\gamma/Z^0$ exchange. In this case, 95\% CL mass limits were placed on HECO production with a cross-section in the range as small as  2~fb;  electric charges from $10e$ to $300e$; and mass limits as high as 2.2~\tev.  In the case of HECOs produced via photon-fusion,  95\% CL lower mass limits were placed on HECOs, with the following properties:  0, spin-\hf, and 1; cross-section reaching down to roughly 1~fb; electric charges in the range $10e$ to $400e$. The corresponding mass limits, that range up to nearly 4~\tev,  are tabulated in Table~\ref{tab:mass-limits-hecos}. 

At the LHC, ATLAS is the only other experiment to have published searches for MMs and HECOs~\cite{ATLAS-13TeV} based on DY production of  MMs or HECOs and, much more recently, on PF and DY production~\cite{Aad:2023}. 

MoEDAL's direct limits on the MM and  HECO mass are the best published to date for, spin-1  MMs with magnetic charge (Q$_{m}$)  1\gd to  10\gd, Spin-0 and Spin-\hf MMs with magnetic charge in the range 2\gd $\le $ Q$_{m}$ $\le$ 10\gd  and HECOs with a charge (Q$_{e}$)  50e $\le $ Q$_{e}$  $\le$ 400e and  10e $< $ Q$_{e}$  $<$ 20e.     
Whereas ATLAS has the best-published mass limits for 1\gd spin-0 and spin-\hf MMs  as well as  HECOs in the range 20e $\le $ Q$_{e}$  $<$ 50e and 2e $\le$ Q$_{e}$  $\le$ 7e.


Importantly, MoEDAL is the sole collider experiment to utilize dedicated detectors able to measure and definitively identify the magnetic charge of the MM  and also to be able to directly calibrate its detectors experimentally for HIPs using heavy ions.

We thank CERN for the  LHC's successful  Run-2 operation and the support staff from our institutions, without whom MoEDAL could not be operated. We acknowledge the invaluable assistance of particular members of the LHCb Collaboration: G.~Wilkinson, R.~Lindner, E.~Thomas and G.~Corti. In addition, we would like to recognize the valuable input from W.-Y.~Song and W.~Taylor of York University on HECO production processes.   Computing support was provided by the GridPP Collaboration, in particular by the Queen Mary University of London and Liverpool grid sites. This work was supported by the UK Science and Technology Facilities Council, via the grants ST/L000326/1, ST/L00044X/1, ST/N00101X/1, ST/T000759/1 and ST/X000753/1; by the Generalitat Valenciana via the projects PROMETEO/2019/087, CIPROM/2021/073 and CIAPOS/2021/88; by Spanish MICIN / AEI / FEDER, EU via the grant PID2021-122134NB-C21; by Spanish MUNI via the mobility grant PRX22/00633; by CSIC via the mobility grant IMOVE23097; by the Physics Department of King's College London; by  NSERC via a project grant; by the V-P Research of the University of Alberta (UofA); by the Provost of the UofA); by UEFISCDI (Romania); by the INFN (Italy);  by a National Science Foundation grant (US) to the University of Alabama MoEDAL group; and, by the National Science Centre, Poland, under research grant 2017/26/E/ST2/00135 and the Grieg grant 2019/34/H/ST2/00707.


\end{document}
